\documentclass[aps,preprint,onecolumn,superscriptaddress]{revtex4}
\usepackage{graphicx,graphics,amssymb,amsmath,subeqnarray}\usepackage{color}

\def\varepsilon{\epsilon}

\begin{document}

\title{Orientational order in concentrated suspensions of spherical microswimmers}

\author{Arthur A. Evans}
\affiliation{Department of Physics, University of California San Diego, 9500 Gilman Drive, La Jolla CA 92093-0354, USA}
\author{Takuji Ishikawa}
\author{Takami Yamaguchi}
\affiliation{Department of Bioengineering and Robotics, Tohoku University, 6-6-01, Aoba, Aramaki, Aoba-ku, 
Sendai 980-8579, Japan}
\author{Eric Lauga}
\affiliation{Department of Mechanical and Aerospace Engineering, University of California San Diego, 9500 Gilman Drive, La Jolla CA 92093-0411, USA}

\date{\today}

\begin{abstract}
We use numerical simulations to probe the dynamics of concentrated suspensions of spherical microswimmers interacting hydrodynamically. Previous work in the dilute limit predicted orientational instabilities of aligned suspensions for both pusher and puller swimmers, which we confirm computationally. Unlike previous work, we show that isotropic suspensions of spherical swimmers are also always unstable. Both types of initial conditions develop long-time polar order, of a nature which depends on the hydrodynamic signature of the swimmer but very weakly on the volume fraction up to very high volume fractions.
\end{abstract}

\maketitle

Living fluids and chemically active colloidal dispersions are modern examples of nonequilibrium hydrodynamic phenomena, presenting tantalizing avenues for both research and industrial application. Recent experiments on motile particles, from collections of microorganisms \cite{DrescherMeasurement,CisnerosExp2007,Dombrowski2004,Riedel2005,Sokolov2007,WuLibchaber2000} to self-propelled colloids \cite{HowseMotileColloids,Fournier2005,Ruckner2007,Paxton2004}, exhibit pattern forming behavior and enhanced transport characteristics that pose  fundamental questions  for the  nonequilibrium statistical mechanics of active systems \cite{ActiveReview}, and have implications
in bio- and nano-engineering \cite{WangReview}. One area of particular interest concerns the hydrodynamics of  microorganisms swimming in  viscous fluids at zero Reynolds number \cite{Lauga:2009p421,BrennenWinetReview,LighthillReview}. 

Several methodologies have been developed in the past to  address the emergence of collective locomotion, corresponding to either microscopic or macroscopic formulations.  Active hydrodynamic equations developed from  non-equilibrium kinetic theory has been the prevailing microscopic approach to the system \cite{SubramanianKoch2009,SaintillanShelley2008,BaskaranMarchetti2009}. By modeling  microswimmers as force-dipoles, these theories build continuum dynamical equations for fields quantifying the long-wavelength properties  of suspensions  of self-propelled particles. The difficulty in dealing with interacting particles leads to the necessity of a dilute assumption, and the lack of any specified structure to the microswimmers (beyond oriented point singularities).

At the other end of the spectrum, thermodynamic models of active media give access to nonlinearities and coupled modes that cannot be derived from a dilute formulation due to relevant (allowed) terms in a dynamical equation \cite{TonerFlock,MishraPatternFormation}.  These ``flocking" models lead to rich dynamics that cannot be captured by the microscopic models, at the expense of introducing phenomenological parameters that may be difficult to explain physically and derive from microscopic considerations \cite{ActiveReview}.

Here we use a  model spherical microswimmer  to address orientational order computationally. With our approach both semi-dilute and concentrated suspensions of spherical swimmers can be considered. Many microorganisms, such as \textit{B. subtilis} or \textit{E. coli} are elongated swimmers, and as a result much work has been done to model the locomotion of rodlike active particles, as well as fabricate synthetic devices of similar aspect ratio; however, colloidal microswimmers, active oil droplets or self-propelled vesicles are likely to be spherical in shape, not to mention the spherical organisms like \textit{Volvox} which appear readily in nature. For all of these reasons it is important to develop an understanding of the hydrodynamic interactions between active spherical particles.  Contrary to predictions from  dilute continuum theories, we find that isotropic suspensions of spherical swimmers  are unstable, and evolve dynamically to take on a long-time state of polar order. This state of polar order, which exists in phenomenological flocking models,  is thus shown here to arise from hydrodynamic interactions.  


\begin{figure}[t]
\includegraphics[width=0.8\textwidth]{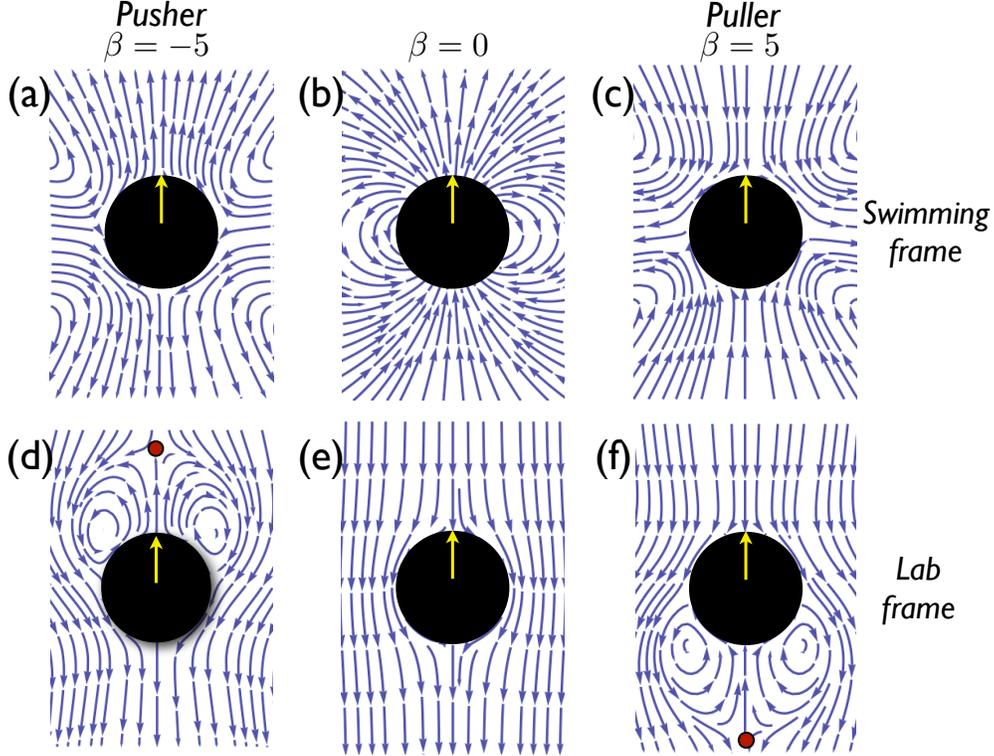}
\caption{\label{swimmers} Flow streamlines of isolated squirmers in the swimming frame (top; a to c) and lab frame (bottom; d to f). 
Left (a and d): Pusher with  a negative stresslet ($\beta=-5$). 
Center (b and e):  Potential flow developed by a squirmer with $\beta=0$. 
Right (c and f): Puller with  a positive stresslet ($\beta=+5$). 
}
\end{figure}

We simulate a system of $N$ spherical swimmers in a cubic box of volume $L^3$, where $L$ is determined from the number of swimmers and preset volume fraction $\phi$, i.e.~$\phi=4\pi N /(3L^3)$ (the radius of the swimmer is  1) and periodic boundary conditions. The swimmer we use, termed a squirmer  \cite{Lighthill1952,Blake1971a},  is a spherical particle that has a prescribed axisymmetric tangential velocity distribution on its surface.  We impose $u_\theta (\theta)=B_1P_1(\cos\theta)+B_2P_2(\cos\theta)$, where $P_n$ is the $n^{th}$ Legendre polynomial, and $\theta=0$ defines the direction  $\b{e}$  in which the squirmer swims, with speed (for a solitary squirmer)  $U\sim B_1$ ($U$ is set to 1). Fluid disturbances in the far field are governed by the ``stresslet" (or force-dipole) of the organisms, quantified by the dimensionless quantity $\beta= B_2/B_1$. The flow field decays as $\sim \beta/r^2$ far from the swimmer, and, in the absence of thermal fluctuations, the stresslet dominates the long-range interactions \cite{Lauga:2009p421}, as illustrated in Fig.~\ref{swimmers}.  Some microorganisms, like the algae  \textit{Chlamydomonas}, generate thrust in front of their bodies pulling themselves therefore through the fluid (``pullers'', with $\beta >0$) while most flagellated cells such the bacteria \textit{E. coli}, or spermatozoa,  generate thrust behind them, and instead are being pushed from the back (``pushers", with $\beta<0$). {Typical values for $\beta$ range from $\approx -1$ for bacteria like \textit{E. Coli} \cite{Drescher2011}, $\approx 0$ for \textit{Volvox} and artificially created squirmers \cite{Short2006,MiuraVesicleMotion2010,ThutupalliSquirmers}, to $\approx +1$ for \textit{Chlamydomonas}} \cite{PedleyChlamy}. 

\begin{figure}[t]
\includegraphics[width=0.5\textwidth]{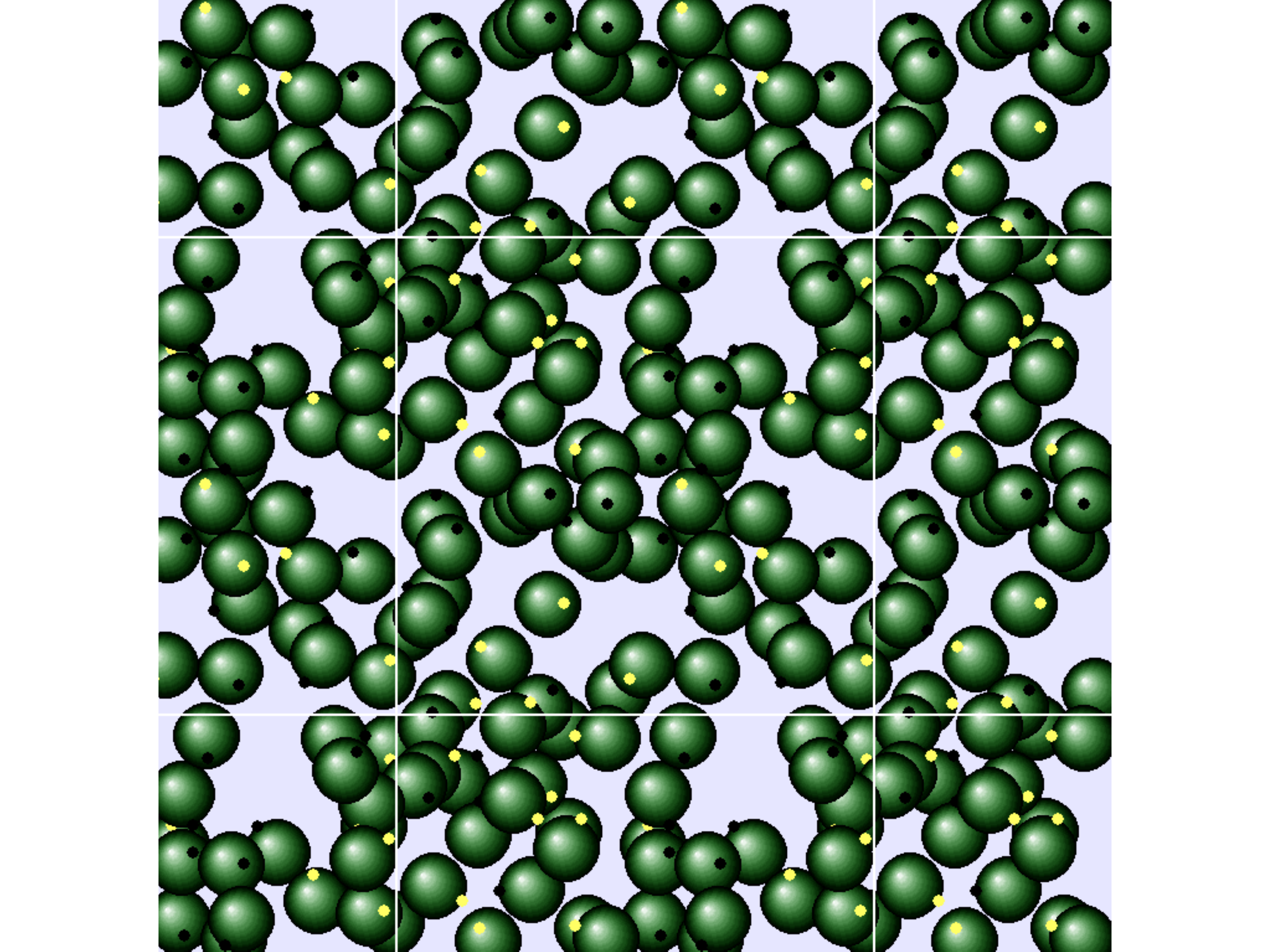}
\caption{\label{snapshot} A snapshot of a simulation with $\phi=0.1$, $\beta=1$ (pullers), and $N=64$. The computational cell is in the middle, with identical copies surrounding this periodic box. Dark dots on the squirmers represent the rear (i.e. $\theta=\pi$), while light dots represent the swimming direction, $\theta=0$.}
\end{figure}

The swimming kinematics of the $N=64$ squirmers are  calculated by enforcing instantaneously the condition of force-  and torque-free swimming. The computational approach is based on Stokesian dynamics, with an analytical treatment of lubrication forces for closely separated swimmers, as well as  short-range repulsive forces to prevent particle overlap, as described in detail in Refs.~\cite{Ishikawa2006,Ishikawa2008}.  A typical snapshot of the simulation is displayed in Fig.~\ref{snapshot}.  The two parameters characterizing the collective swimming dynamics are thus the swimmer volume fraction,  $\phi$,  and the swimmer stresslet, $\beta$. In order to probe the development of order in our system,  we define a polar order parameter, $P$, based on the orientation vector $\b{e}$ of the particles, namely $P(t)=|\sum_i^N\b{e}_i(t)|/N$. If every particle is swimming in the same direction (polar order) then $P=1$, while for isotropic orientation we expect $P\sim1/\sqrt{N}$.

\begin{figure}[t!]
\includegraphics[width=0.55\textwidth]{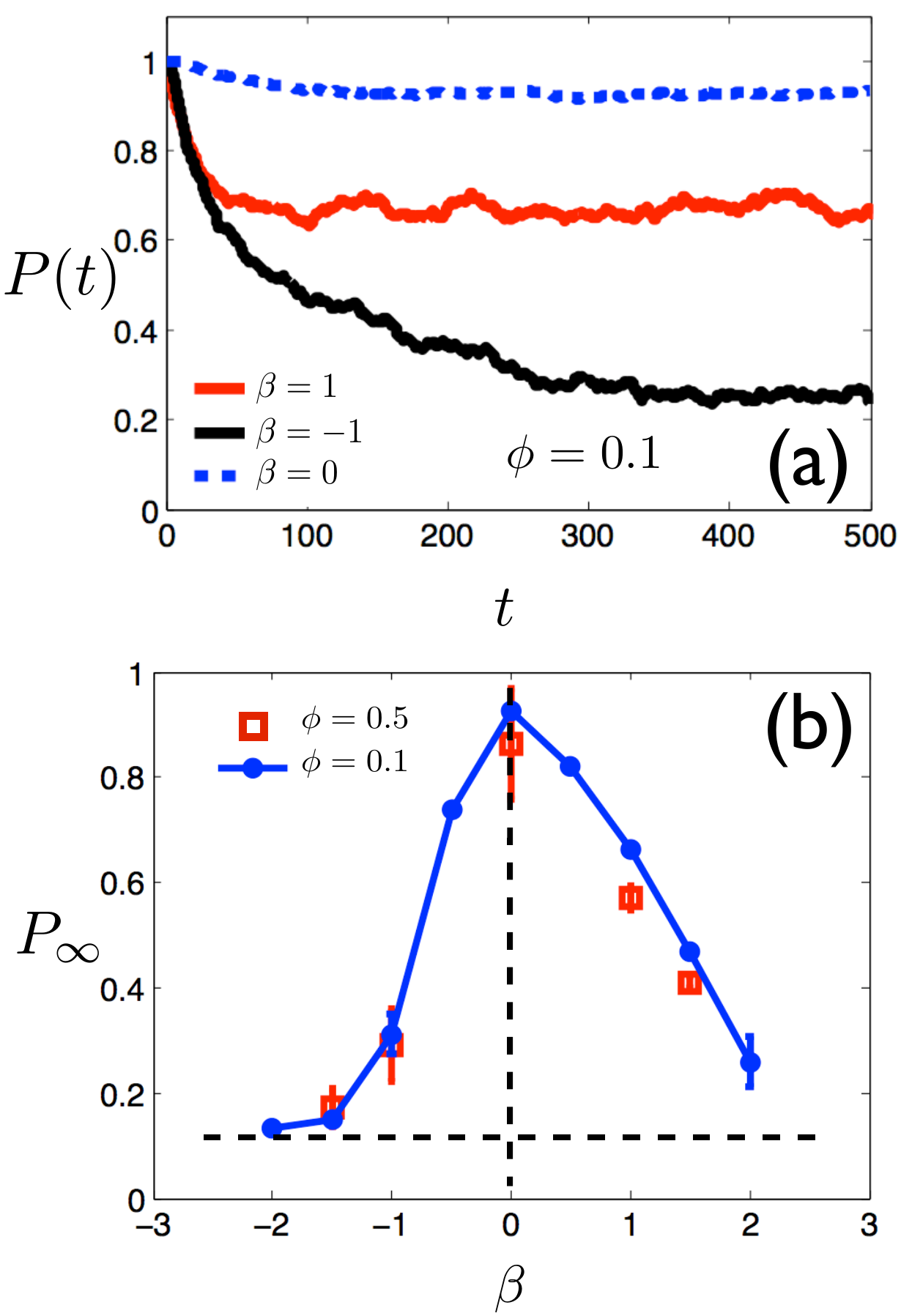}
\caption{\label{examples} Global order of semi-dilute suspensions of pushers and pullers starting from aligned state. 
(a): Order parameter, $P(t)$,  for pushers ($\beta=-1$), pullers ($\beta=1$), and potential swimmers ($\beta=0$).  Each simulation starts from an initially aligned state (with random positions) and decays to a finite order, $P_\infty$, after a characteristic decay time that depends on the stresslet coefficient $\beta$.  
(b): Long-time order parameter, $P_\infty$, for initially aligned suspensions. Each data point represents an average over an ensemble (five separate realizations) and time (performed after initial transient). Error bars represent standard deviations of the ensemble averaging process. The horizontal dashed line indicates the expected result for isotropic suspensions ($N=64$).}
\end{figure}

In the dilute limit, continuum theories for slender self-propelled rods have predicted that aligned suspensions are always unstable, for both pushers and pullers.  On the contrary, isotropically oriented states are only unstable for pushers  ($\beta<0$) while pullers remain in an isotropic state \cite{SaintillanShelley2008}. The physical nature of the isotropic instability is described as resulting from long-range hydrodynamic extensional disturbances that cause reorientation of anisotropic particles, for which spherical swimmers are immune. In the dilute limit, no instability of isotropic suspensions due to long-range hydrodynamic interactions is thus expected to occur for spherical particles \cite{SaintillanShelley2008,SubramanianKoch2009}.

In Fig.~\ref{examples}a we plot the time-evolution of the polar order parameter, $P(t)$, for suspensions starting in the aligned state with three different stresslet values, $\beta=-1$ (pusher), 0 (potential swimmer) and $+1$ (puller).  The initial positions of the swimmers are taken to be random.  Over the semi-dilute to concentrated  range of volume fractions, $\phi=0.1 - 0.5$ (the results in Fig.~\ref{examples}a are shown for  $\phi=0.1$),  we observe the system to systematically decay from perfect order ($P=1$) to some finite long-time value ($0<P_\infty <1$).   The decay time over which the suspension is driven to this new ordered state  depends on the stresslet value $\beta$, and larger values of the stresslet disturb the fluid more violently, causing reorientation, and loss of polar order, more quickly; $\beta$ can thus be interpreted as the  speed at which orientation decorrelation propagates throughout the suspension.

To further characterize long-time global order, we perform ensemble averages on five realizations of our simulations, starting with an aligned orientations  and  random  positions, for volume fractions in the range $\phi=0.1$ to $0.5$ and stresslets varying from $\beta=-2$ to $+2$. Upon doing ensemble averages, we define  a long-time order parameter $P_\infty$ by averaging over the time period after the initial decay from alignment until the end of the simulation; the results are shown in Fig.~\ref{examples}b. 
In the range of volume fractions studied, the long-time order is  essentially independent of the value of $\phi$, but  strongly depends on the value of the stresslet $\beta$.  Larger values of $|\beta|$ lead to increased swimmer-swimmer reorientations due hydrodynamic interactions, and thus lead to decreased values of $P_\infty$. In addition, we observe a marked asymmetry between pushers and pullers, and for a given value of $|\beta|$, pullers are systematically  more ordered. For large values of $|\beta|$, the global order disappears, as shown by the dashed line in Fig. \ref{examples}b denoting the $1/\sqrt{N}$ isotropy that is expected in our system. This is in contrast with analytical predictions in the dilute limit where suspensions of pullers are expected to be systematically driven to orientational isotropy \cite{SaintillanShelley2008}.

 \begin{figure}[t!]
\includegraphics[width=0.5\textwidth]{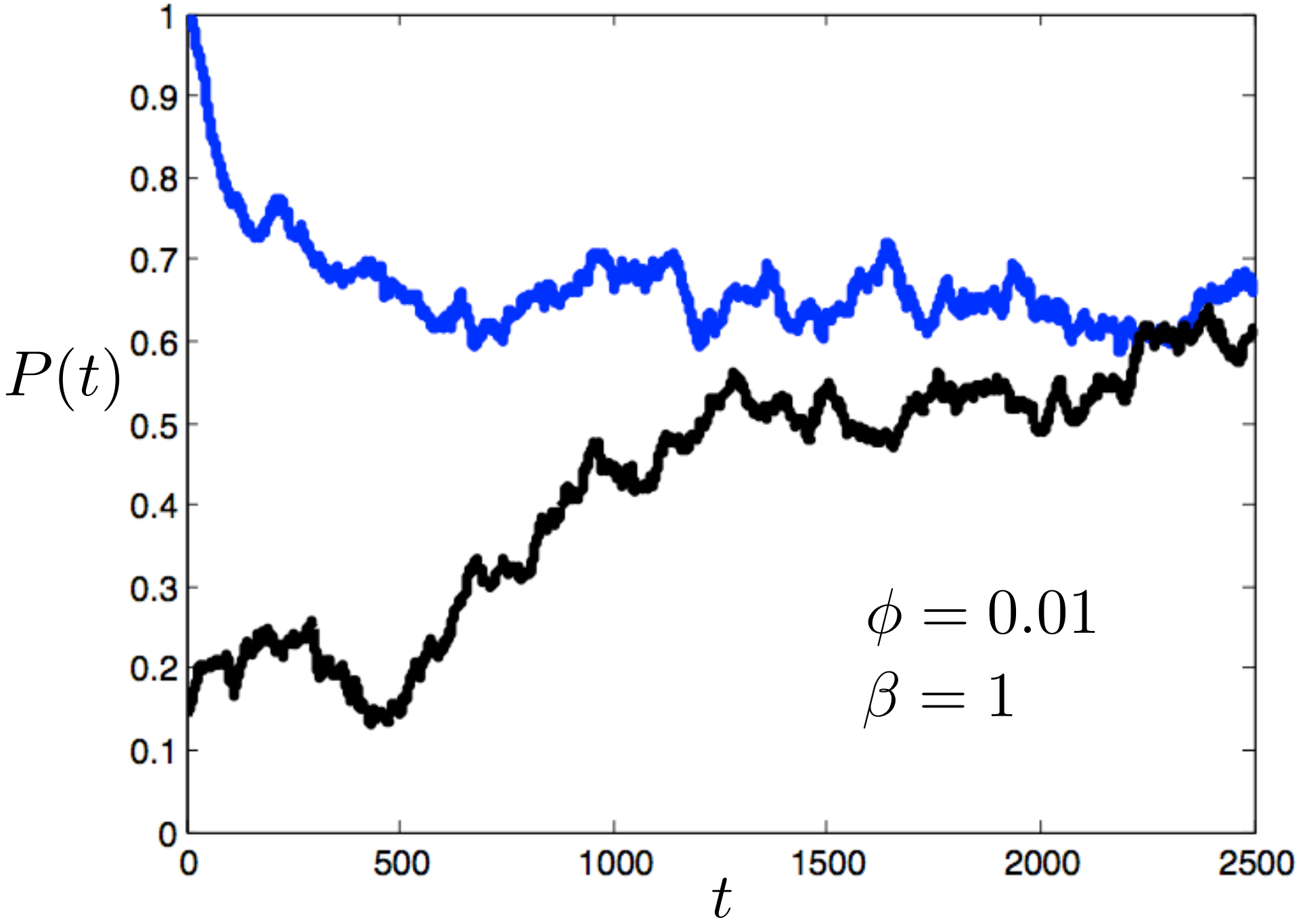}
\caption{\label{isoalign} Time-evolution of the order parameter, $P(t)$, for  initially aligned (top) and isotropic (bottom) suspensions (in both cases  initial positions are random). 
The case shown is for a puller swimmer ($\beta=1$) with a volume fraction $\phi=0.01$. Both aligned and isotropic states are unstable and approach similar finale state of global orientational order at long times.}
\end{figure}

It has been suggested in the past that near-field interactions are the dominant mechanism for polar ordering \cite{Ishikawa2008}. To address the pusher-puller asymmetry, we take a detailed look at the flow field near the swimmers.  The far-field difference between a pusher and a puller leads in the near-field to a change in the location of the stagnation point, see Fig.~\ref{swimmers}: for a pusher this stagnation point leads the swimmer, while for a puller it trails it. This asymmetry causes the different collisional situations (head to head, head to tail, and tail to tail) to produce different reorientations between pushers and pullers. From previous numerical work, it is known that  head-to-head interactions are far more likely to occur \cite{IshikawaHota2006}. For pushers the head-to-head orientation is stable to rotation (see Fig.~\ref{swimmers}d): the presence of a stagnation point establishes vortices near the surface of the swimmer, and in a head to head collision both pairs of vortices interact in a manner as  to maintain this orientation. Pullers, on the other hand, are unstable in this configuration, as their vortices trail in such a collision.  In terms of contributing to order, head to head orientations yields $P\sim0$, and thus any instability of this configuration, as expected for pullers, will lead to an  increase of polar order, hence the asymmetry between pushers and pullers seen in Fig.~\ref{examples}b.


In contrast to the aligned case, orientation instabilities in isotropic suspensions was predicted by continuum theories to exist only for pushers, and only then when the swimmers have a nonspherical shape  \cite{SaintillanShelley2008}.  
We performed numerical simulations to probe the long-time behavior of initially isotropic suspensions. We show in Fig.~\ref{isoalign} the evolution in time of the order parameter for  suspensions of pullers ($\beta=1$) at very low volume fraction ($\phi=0.01$), for both aligned (top) and isotropic (bottom) initial conditions. The results for pushers are similar. We see not only that  both sets of initial conditions are unstable, but  also that both are driven at long times to a similar, and intermediate, value of the order parameter.  The time scale over which the instability takes place is long compared to the case of more concentrated suspensions, and thus it is plausible that in the limit $v\rightarrow 0$ the instability would disappear, allowing us to reconcile our numerical results with those of continuum theories.

\begin{figure}[t!]
\includegraphics[width=0.55\textwidth]{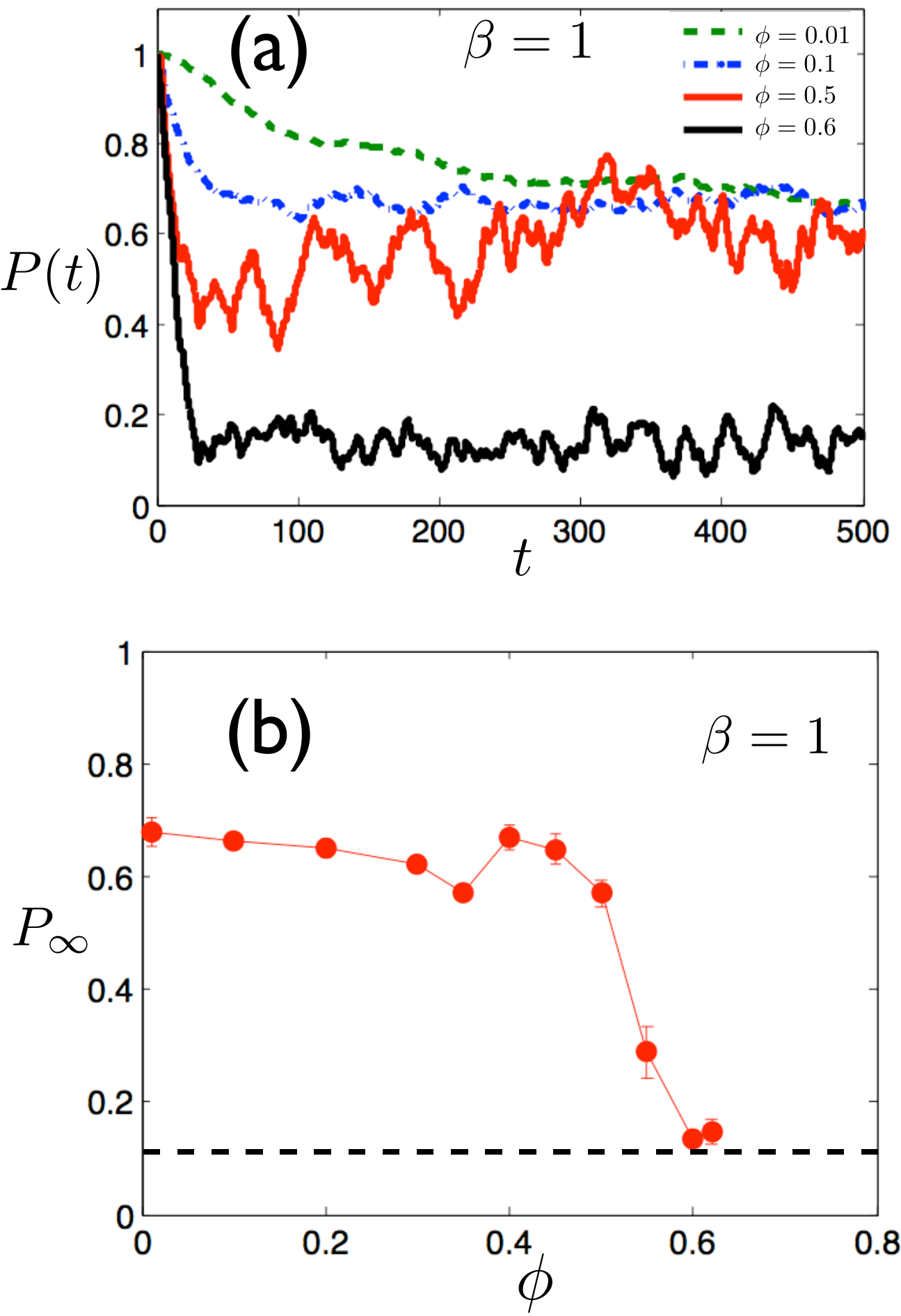}
\caption{\label{Pvsv} Polar order as function of volume fraction. (a) Samples of order decay for initially-aligned pullers ($\beta=1$); (b) Long-time order as a function of volume fraction, for pullers ($\beta=1$). For semi-dilute suspensions  there is only a weak dependence on the volume fraction.  At larger volume fractions the order drops sharply to isotropy (indicated by dashed line).  Error bars show standard deviation over many realizations.}
\end{figure}

\begin{figure*}[t!]
\includegraphics[width=0.75\textwidth]{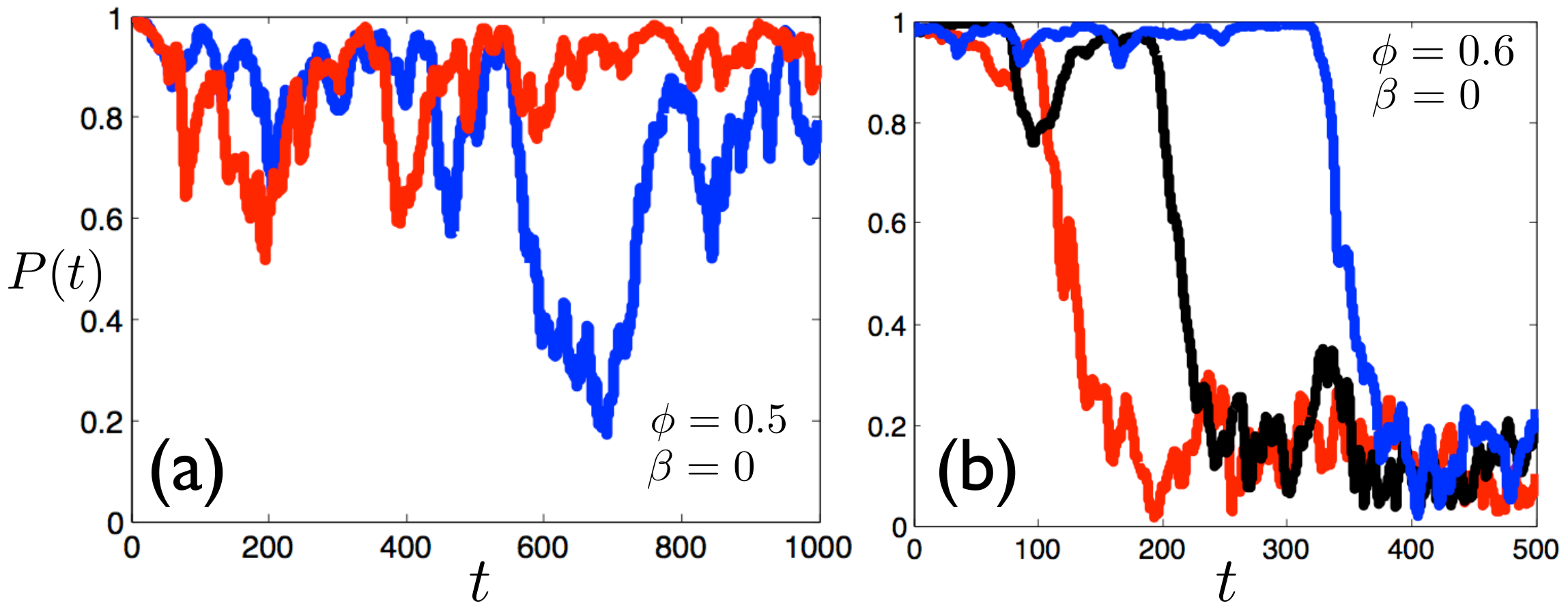}
\caption{\label{transient} Behavior at high volume fraction for $\beta=0$. (a): Dynamics starting from two different initial positions for $\phi=0.5$;  the fluctuations about the long time average  $P_\infty$ become  large at high volume fractions; (b): Dynamics starting from three different initial positions for $\phi=0.6$;  above $\phi\sim 0.5$, the transition to isotropy can show significant delay times, likely associated with an inability to sample all of phase space.}
\end{figure*}

How does increasing the volume fraction affect this order?   
We show in Fig.~\ref{Pvsv}a the evolution in time of the order parameter for different volume fractions, $\phi$, of pullers ($\beta=1$).  The averaged value of the order parameter, $P_\infty$, is plotted as a function of the volume fraction in Fig.~\ref{Pvsv}b.  We see that as  the volume fraction starts increasing away from the dilute limit, the average value of the order parameter  changes only slightly over a wide range of volume fractions. When the volume fraction reaches   $\phi\sim0.5$,  fluctuations in the order parameter are observed to become  larger, until finally the order disappears near $\phi\sim 0.6$.  Suspensions of pushers display a similar dependence, although with a decreased overall magnitude of the order parameter due to the pusher-puller asymmetry noted earlier. 
The fluctuating behavior observed at high concentrations is further illustrated in Fig.~\ref{transient}a. We also show in Fig.~\ref{transient}b how different initial conditions for the same dense suspension  can yield drastically different results, indicating that particles are ``trapped" in their initial conditions before finally escaping to the isotropic steady state.

In summary, we have used in this paper numerical simulations to address the instabilities and long-time order of semi-dilute and dense suspensions of spherical swimmers. We have shown that spherical squirmers, starting from either an aligned or isotropic state, develop long-time polar order due to hydrodynamic interactions, of a nature which depends on the hydrodynamic signature of the swimmer but very weakly on the volume fraction up to the dense regime.  Our results show thus non-trivial differences with dilute, dipole continuum models, but display similarities  with phenomenological flocking models. 

This work was funded in part by the National Science Foundation through grant CBET-0746285 (EL) and a 2010 East Asia and Pacific Summer Institute Research Fellowship (AAE).

\end{document}